\documentclass[acmsmall,nonacm]{acmart}
\setcopyright{none}

\usepackage[capitalise]{cleveref}
\usepackage{microtype}

\usepackage[frozencache,cachedir=.]{minted}
\usepackage{mdframed}
\BeforeBeginEnvironment{minted}{\begin{mdframed}[linewidth=0pt]\medskip}
\AfterEndEnvironment{minted}{\medskip\end{mdframed}}

\usepackage{forest}
\useforestlibrary{linguistics}
\usepackage{xcolor}
\usepackage{tikz}
\usepackage{yquant}

\newcommand{\strue}{\texttt{\#t}}
\newcommand{\sfalse}{\texttt{\#f}}
\newcommand{\rkt}[1]{\mintinline{racket}{#1}}
\newcommand{\h}[1]{\mintinline{racket}{#1}}
\newcommand{\xgate}{\textsc{x}}
\newcommand{\cx}{\textsc{cx}}
\newcommand{\ccx}{\textsc{ccx}}
\newcommand{\had}{\textsc{h}}
\newcommand{\ket}[1]{|#1\rangle}

\begin{document}

\title{Scheme Pearl: Quantum Continuations}

\author{Vikraman Choudhury}
\orcid{0000-0003-2030-8056}
\affiliation{
  \department{School of Computing Science}
  \institution{University of Glasgow}
  \city{Glasgow}
  \postcode{G12 8QQ}
  \country{UK}
}
\email{vikraman.choudhury@glasgow.ac.uk}

\author{Borislav Agapiev}
\email{borislav.agapiev@gmail.com}

\author{Amr Sabry}
\orcid{0000-0002-1025-7331}
\affiliation{
  \department{Department of Computer Science}
  \institution{Indiana University}
  \city{Bloomington}
  \postcode{47408}
  \country{USA}
}
\email{sabry@indiana.edu}

\begin{abstract}
  We advance the thesis that the simulation of quantum circuits is fundamentally
  about the efficient management of a large (potentially exponential) number of
  delimited continuations. The family of Scheme languages, with its efficient
  implementations of first-class continuations and with its imperative
  constructs, provides an elegant host for modeling and simulating quantum
  circuits.
\end{abstract}

\maketitle

\section{Introduction}


The non-intuitive properties and widely-heralded computational advantage of
quantum computing are often associated with concepts like superposition,
entanglement, and complementarity~\cite{khrennikov}. For the programming
language aficionados, we instead present a different perspective: the power of
quantum computing can be attributed to an extraordinarily efficient management
of an exponential number of continuations.

We demonstrate that quantum computing is just one step away from
well-established techniques with roots in the Scheme community. Specifically,
quantum computing can be reduced to the use of continuations to implement
backtracking search and
non-determinism~\cite{10.11451086365.1086390,kiselyovmusabry2021,10.1145/351240.351258,10.11451016850.1016861},
augmented with a global mechanism to manage these continuations. Thus, in
addition to its expository role, this paper opens up a new perspective where
clever implementation techniques of continuations~\cite{10.1145/93542.93554},
especially those relying on speculative evaluation, can enable more efficient
classical simulations of quantum algorithms in particular cases.

\subsubsection*{Outline}

We begin the next section with a review of the aspects of quantum computing that
are necessary for our exposition. We focus on a minimal model of quantum
circuits that is still rich enough to express general quantum computations up to
arbitrary small errors, expressed inside Racket.

To make the connections to continuations apparent, in~\cref{sec:three}, we
express the semantics of these quantum circuits using a slight generalization of
search trees where the edges are decorated with probability amplitudes. These
amplitudes may be positive or negative to express constructive or destructive
interference patterns.

In~\cref{sec:four}, we give an interpreter that uses delimited continuations to
evaluate the quantum circuits. The interpreter is parameterized by a function
that collects and manages the generated continuations. A management strategy
that invokes all the continuations and appends their results produces full
information about the probabilistic behavior of the quantum computation.
Although instructive, this information requires an exponential cost to generate
and outputs information that is physically unobservable.

An adjustment to the previous strategy, given in~\cref{sec:five}, that produces
physically observable results is to calculate interference patterns among the
continuations allowing for some branches to cancel each other, and then sample
from the remaining branches. The straightforward implementation of this latter
strategy would still be exponential, but it might be possible in some cases to
discover more ingenious implementation strategies, which is an idea we leave for
consideration in future work, as explained in the concluding section.

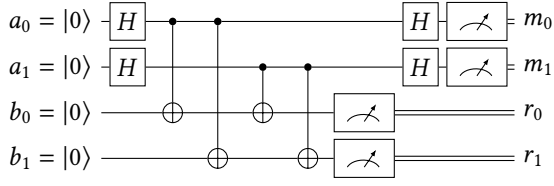
\begin{figure}[t]
  \begin{center}
    \begin{tikzpicture}[scale=1.0,every label/.style={rotate=40, anchor=south west}]
      \begin{yquant*}[operators/every barrier/.append style={red, thick}]
        qubit {$a_0=\ket0$} a0;
        qubit {$a_1=\ket0$} a1;
        qubit {$b_0=\ket0$} b0;
        qubit {$b_1=\ket0$} b1;
        box {$H$} a0;
        box {$H$} a1;
        cnot b0 | a0;
        cnot b1 | a0;
        cnot b0 | a1;
        cnot b1 | a1;
        align -;
        measure b0;
        measure b1;
        align -;
        box {$H$} a0;
        box {$H$} a1;
        align -;
        measure a0;
        measure a1;
        output {$m_0$} a0;
        output {$m_1$} a1;
        output {$r_0$} b0;
        output {$r_1$} b1;
      \end{yquant*}
    \end{tikzpicture}
  \end{center}
  \caption{\label{fig:simon} Circuit for a small instance of Simon's
    problem with $n=2$ and $a=3$.  Given a 2-1 function
    $f : \mathbb{B}^n  \rightarrow \mathbb{B}^n$
    with the property that there exists an $a$
    such that $f(x) = f(x~\textsc{xor}~a)$ for all $x$, the problem is
    to determine $a$.}
\end{figure}

\section{Quantum Circuits}
\label{sec:two}

The circuit model is a well-established universal model of quantum
computing~\cite{nielsen_chuang_2010}. We introduce the main idea using
an extended example and then define a small circuit language whose
semantics we model using continuations.

\subsection{An Example: Simon's Problem}

As shown in~\cref{fig:simon}, a typical quantum circuit consists of an
initialization phase (on the left), followed by sequential and parallel
compositions of elementary gates, and ending with measurements. Each horizontal
line in the circuit corresponds to a quantum bit (qubit). The initial state, and
the result of measuring each qubit, is a classical bit (a boolean value). As we
explain below, between initialization and measurement, the state of the qubit
can be in a \emph{superposition} of boolean values, or even be \emph{entangled}
with other qubits.

The initial state of each of the four qubits in~\cref{fig:simon} is $\ket{0}$,
which is essentially equivalent to a classical boolean 0. In preparation for
possible entanglement among the four qubits, we collect the individual values
into an aggregate $\ket{0000}$, where, by convention, the most significant
position refers to the top wire. The first step in the execution is to apply the
Hadamard ($H$) gate. The gate maps $\ket{0}$ to $\frac{1}{\sqrt{2}} (\ket{0} +
  \ket{1})$ and $\ket{1}$ to $\frac{1}{\sqrt{2}} (\ket{0} - \ket{1})$. Both
results of applying $H$ represent states that are in equal superpositions of
$\ket{0}$ and $\ket{1}$, with the sign expressing whether the values will lead
to constructive or destructive interference. The aggregate state is now:

\[
  \frac{1}{2} ~ \left( (\ket{0}+\ket{1})~(\ket{0}+\ket{1})~\ket{00} \right) =
  \frac{1}{2} ~(\ket{0000} + \ket{0100} + \ket{1000} + \ket{1100})
\]

The next four gates are all \emph{controlled-not} gates which negate the target
bit (marked with $\oplus$) if the control bit is true. The execution therefore
continues with the following states:

\[\begin{array}{rcl}
    \frac{1}{2} ~(\ket{0000} + \ket{0100} + \ket{1000} + \ket{1100})
     & \rightarrow &
    \frac{1}{2} ~(\ket{0000} + \ket{0100} + \ket{1010} + \ket{1110}) \\[2ex]
     & \rightarrow &
    \frac{1}{2} ~(\ket{0000} + \ket{0100} + \ket{1011} + \ket{1111}) \\[2ex]
     & \rightarrow &
    \frac{1}{2} ~(\ket{0000} + \ket{0110} + \ket{1011} + \ket{1101}) \\[2ex]
     & \rightarrow &
    \frac{1}{2} ~(\ket{0000} + \ket{0111} + \ket{1011} + \ket{1100})
  \end{array}\]

At this point, the two qubits in the least significant positions are measured.
By inspection, we see that there is an equal probability of these two qubits
being 00 or 11. Without any loss of generality, we assume these qubits are
measured as 11. Then, the entire state collapses to:
\[
  \frac{1}{\sqrt{2}}~(\ket{0111} + \ket{1011})
\]
and ignoring the now irrelevant measured qubits, the top two qubits are in the
state:
\[
  \frac{1}{\sqrt{2}}~(\ket{01} + \ket{10})
\]
Applying the two $H$ gates, the state before measurement is:
\[
  \frac{1}{2\sqrt{2}}~(\ket{00} - \ket{01} + \ket{10} - \ket{11} +
  \ket{00} + \ket{01} - \ket{10} - \ket{11})
\]
The values with the same sign interfere constructively and the values
with opposite signs interfere destructively leading to the simplified
state:
\[
  \frac{1}{2\sqrt{2}}~(2 \ket{00} - 2 \ket{11}) =
  \frac{1}{\sqrt{2}}~(\ket{00} - \ket{11})
\]
A measurement of the two qubits is equally likely to produce 00 or 11 as the
value of the hidden~$a$. In the former case, the algorithm will have produced
the vacuous identity that $f(x) = f(x)$ and the circuit needs to be re-executed.
In the latter case, the algorithm successfully terminates with the correct
decimal value 3 for $a$.

\subsection{An Approximately Universal Circuit Language}

For theoretical studies of quantum computing, it is particularly appealing to
exploit two facts which justify focus on a tiny foundational core. First, by the
principle of \emph{deferred measurement}~\cite{nielsen_chuang_2010}, it is
always possible to defer all measurements to the last step of a quantum
computation. This allows one to isolate what is called ``pure quantum
computing'' which consists of sequential and parallel compositions of elementary
gates not involving measurements. Second, if one is willing to tolerate
arbitrary small errors, then all of pure quantum computing just needs two
elementary gates~\cite{aharonov:toffolihadamard}: the Hadamard gate introduced
above and the \emph{controlled-controlled-not} (or Toffoli) gate which negates
its target qubit when \emph{both} control qubits are true.

We therefore restrict our attention to circuits built from sequences of the
following two elementary operations:

\begin{minted}{racket}
(define-syntax-rule  (CCX a b c)  `(ccx ,a ,b ,c))
(define-syntax-rule  (H a)        `(h ,a))
\end{minted}
where \rkt{a}, \rkt{b}, and \rkt{c} are natural numbers giving the index of the
qubit in question, or boolean constants \strue\ or \sfalse. As is obvious,
\rkt{CCX} denotes the controlled-controlled-not gate, and \rkt{H} denotes the
Hadamard gate. For convenience, we further introduce the following two
abbreviations:
\begin{minted}{racket}
(define-syntax-rule  (X a)        (CCX #t #t a))
(define-syntax-rule  (CX a b)     (CCX #t a b))
\end{minted}
which introduce special cases of \rkt{CCX} where one or both control qubits are
known to be true. Here \rkt{X} denotes the not gate, and \rkt{CX} denotes the
controlled-not gate. For example, the pure part of the circuit
in~\cref{fig:simon} can be written as the following expression:

\begin{minted}{racket}
  (list
    (H 0)
    (H 1)
    (CX 0 2)
    (CX 0 3)
    (CX 1 2)
    (CX 1 3)
    (H 0)
    (H 1))
\end{minted}

\section{Continuations and Search Trees}
\label{sec:three}

Before considering quantum circuits, let us examine a toy example of an
arithmetic expression. We illustrate the use of continuations to evaluate~$x +
(y * 5)$ where~$x$ can take the value 1 or 2, and~$y$ can take the value 3 or 4.
We begin by visualizing the evaluation as a tree:

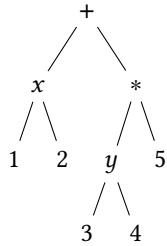
\begin{center}
  \begin{forest}
    [$+$
      [$x$ [1] [2]]
        [$*$ [$y$ [3] [4]] [5]]
    ]
  \end{forest}
\end{center}

Given primitives \h{shift} and \h{reset} to manipulate
continuations~\cite{10.1145/91556.91622}, the four possible results of the
expression can be collected in a list as follows.  We set a continuation
delimiter~\cite{10.1145/73560.73576} at the beginning of the evaluation (i.e.,
at the root of the tree), and then every time we encounter a node with a choice,
we capture and \emph{shift} two copies of the continuation to explore both
branches. Each individual result is wrapped in a list and the results of
invoking the continuations are appended:

\begin{minted}{racket}
(define (choose^ a b)
  (shift k (append (k a) (k b))))

(define expr
  (reset (let ([x (choose^ 1 2)]
               [y (choose^ 3 4)])
           (list (+ x (* 5 y))))))
\end{minted}
\noindent The result of evaluating \h{expr} is a list.
\begin{minted}{racket}
> expr
'(16 21 17 22)
\end{minted}

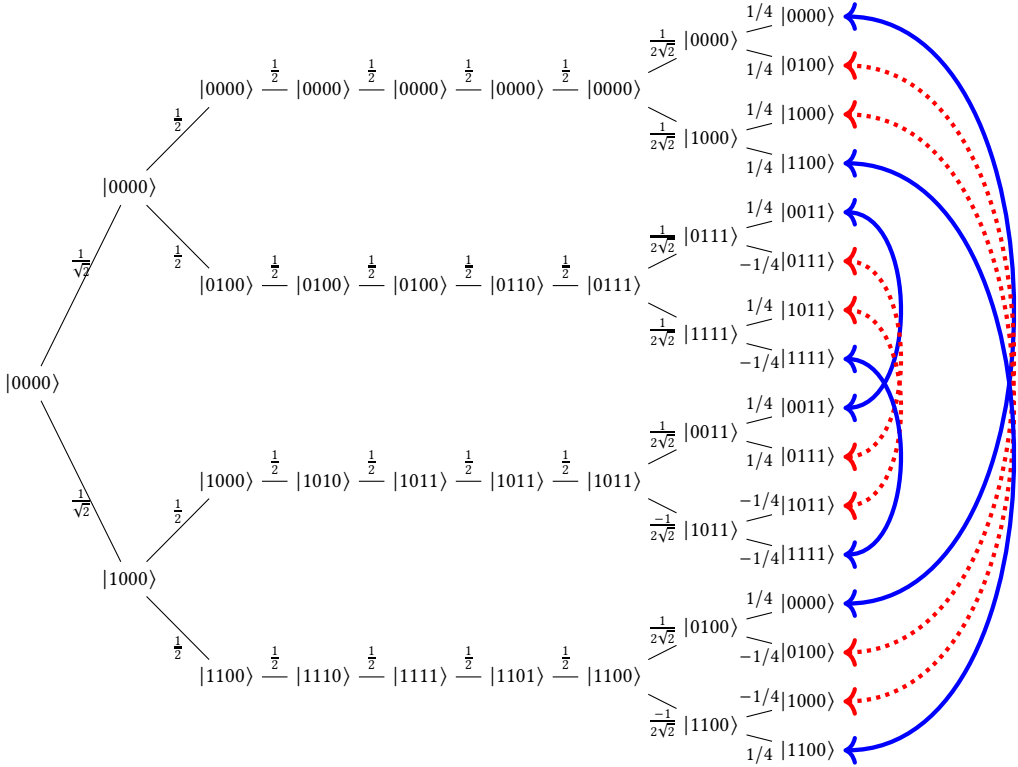
\begin{figure}[t]
  \begin{center}
    {\footnotesize
      \begin{forest}
        for tree={grow'=0}
        [ $\ket{0000}$
        [ $\ket{0000}$, edge label={node[midway,above,font=\scriptsize]{$\frac{1}{\sqrt{2}}$}}
              [$\ket{0000}$, edge label={node[midway,above,font=\scriptsize]{$\frac{1}{2}$}}
                  [$\ket{0000}$, edge label={node[midway,above,font=\scriptsize]{$\frac{1}{2}$}}
                      [$\ket{0000}$, edge label={node[midway,above,font=\scriptsize]{$\frac{1}{2}$}}
                          [$\ket{0000}$, edge label={node[midway,above,font=\scriptsize]{$\frac{1}{2}$}}
                              [$\ket{0000}$, edge label={node[midway,above,font=\scriptsize]{$\frac{1}{2}$}}
                                  [$\ket{0000}$, edge label={node[midway,above,font=\scriptsize]{$\frac{1}{2\sqrt{2}}$}}
                                      [$\ket{0000}$, name=a00, edge label={node[midway,above,font=\scriptsize]{$1/4$}}]
                                      [$\ket{0100}$, name=a01, edge label={node[midway,below,font=\scriptsize]{$1/4$}}]
                                  ]
                                  [$\ket{1000}$, edge label={node[midway,below,font=\scriptsize]{$\frac{1}{2\sqrt{2}}$}}
                                      [$\ket{1000}$, name=a02, edge label={node[midway,above,font=\scriptsize]{$1/4$}}]
                                      [$\ket{1100}$, name=a03, edge label={node[midway,below,font=\scriptsize]{$1/4$}}]
                                  ]
                              ]]]]]
              [$\ket{0100}$, edge label={node[midway,below,font=\scriptsize]{$\frac{1}{2}$}}
                  [$\ket{0100}$, edge label={node[midway,above,font=\scriptsize]{$\frac{1}{2}$}}
                      [$\ket{0100}$, edge label={node[midway,above,font=\scriptsize]{$\frac{1}{2}$}}
                          [$\ket{0110}$, edge label={node[midway,above,font=\scriptsize]{$\frac{1}{2}$}}
                              [$\ket{0111}$, edge label={node[midway,above,font=\scriptsize]{$\frac{1}{2}$}}
                                  [$\ket{0111}$, edge label={node[midway,above,font=\scriptsize]{$\frac{1}{2\sqrt{2}}$}}
                                      [$\ket{0011}$, name=a04, edge label={node[midway,above,font=\scriptsize]{$1/4$}}]
                                      [$\ket{0111}$, name=a05, edge label={node[midway,below,font=\scriptsize]{$-1/4$}}]
                                  ]
                                  [$\ket{1111}$, edge label={node[midway,below,font=\scriptsize]{$\frac{1}{2\sqrt{2}}$}}
                                      [$\ket{1011}$, name=a06, edge label={node[midway,above,font=\scriptsize]{$1/4$}}]
                                      [$\ket{1111}$, name=a07, edge label={node[midway,below,font=\scriptsize]{$-1/4$}}]
                                  ]
                              ]]]]]
          ]
          [ $\ket{1000}$, edge label={node[midway,below,font=\scriptsize]{$\frac{1}{\sqrt{2}}$}}
              [$\ket{1000}$, edge label={node[midway,above,font=\scriptsize]{$\frac{1}{2}$}}
                  [$\ket{1010}$, edge label={node[midway,above,font=\scriptsize]{$\frac{1}{2}$}}
                      [$\ket{1011}$, edge label={node[midway,above,font=\scriptsize]{$\frac{1}{2}$}}
                          [$\ket{1011}$, edge label={node[midway,above,font=\scriptsize]{$\frac{1}{2}$}}
                              [$\ket{1011}$, edge label={node[midway,above,font=\scriptsize]{$\frac{1}{2}$}}
                                  [$\ket{0011}$, edge label={node[midway,above,font=\scriptsize]{$\frac{1}{2\sqrt{2}}$}}
                                      [$\ket{0011}$, name=a08, edge label={node[midway,above,font=\scriptsize]{$1/4$}}]
                                      [$\ket{0111}$, name=a09, edge label={node[midway,below,font=\scriptsize]{$1/4$}}]
                                  ]
                                  [$\ket{1011}$, edge label={node[midway,below,font=\scriptsize]{$\frac{-1}{2\sqrt{2}}$}}
                                      [$\ket{1011}$, name=a10, edge label={node[midway,above,font=\scriptsize]{$-1/4$}}]
                                      [$\ket{1111}$, name=a11, edge label={node[midway,below,font=\scriptsize]{$-1/4$}}]
                                  ]
                              ]]]]]
              [$\ket{1100}$, edge label={node[midway,below,font=\scriptsize]{$\frac{1}{2}$}}
                  [$\ket{1110}$, edge label={node[midway,above,font=\scriptsize]{$\frac{1}{2}$}}
                      [$\ket{1111}$, edge label={node[midway,above,font=\scriptsize]{$\frac{1}{2}$}}
                          [$\ket{1101}$, edge label={node[midway,above,font=\scriptsize]{$\frac{1}{2}$}}
                              [$\ket{1100}$, edge label={node[midway,above,font=\scriptsize]{$\frac{1}{2}$}}
                                  [$\ket{0100}$, edge label={node[midway,above,font=\scriptsize]{$\frac{1}{2\sqrt{2}}$}}
                                      [$\ket{0000}$, name=a12, edge label={node[midway,above,font=\scriptsize]{$1/4$}}]
                                      [$\ket{0100}$, name=a13, edge label={node[midway,below,font=\scriptsize]{$-1/4$}}]
                                  ]
                                  [$\ket{1100}$, edge label={node[midway,below,font=\scriptsize]{$\frac{-1}{2\sqrt{2}}$}}
                                      [$\ket{1000}$, name=a14, edge label={node[midway,above,font=\scriptsize]{$-1/4$}}]
                                      [$\ket{1100}$, name=a15, edge label={node[midway,below,font=\scriptsize]{$1/4$}}]
                                  ]
                              ]]]]]
          ]
        ]
        \draw[<->,ultra thick,blue] (a00) to [out= east,in= east] (a12);
        \draw[<->,ultra thick,red,dotted] (a01) to [out= east,in= east] (a13);
        \draw[<->,ultra thick,red,dotted] (a02) to [out= east,in= east] (a14);
        \draw[<->,ultra thick,blue] (a03) to [out= east,in= east] (a15);
        \draw[<->,ultra thick,blue] (a04) to [out= east,in= east] (a08);
        \draw[<->,ultra thick,red,dotted] (a05) to [out= east,in= east] (a09);
        \draw[<->,ultra thick,red,dotted] (a06) to [out= east,in= east] (a10);
        \draw[<->,ultra thick,blue] (a07) to [out= east,in= east] (a11);
      \end{forest}
    }
  \end{center}
  \caption{\label{fig:tree}Visualizing the evaluation of the pure gates in~\cref{fig:simon}}
\end{figure}

We now adapt this idea to simulate the quantum circuit in~\cref{fig:simon}. We
begin by visualizing the evaluation of the pure gates as a tree, similar to the
arithmetic expression above. The resulting tree, shown in~\cref{fig:tree},
starts with the initial state at the root. The semantics of \rkt{X}, \rkt{CX},
and \rkt{CCX} is straightforward, only making a local change to the state. At
each occurrence of a Hadamard gate, evaluation splits into two branches. The
crucial aspect of the tree is that the edges are labeled with probability
amplitudes that are cumulatively multiplied as evaluation progresses. These
amplitudes capture dependencies among the different paths that are highlighted
with the colored connectors on the right. Each dashed red arrow connects two
identical states reached by different execution paths with opposite probability
amplitudes: they annihilate. The solid blue arrows similarly connect two
identical states reached by different execution paths, but with probability
amplitudes of the same sign: they reinforce each other. Thus, the final state of
the execution is~$\frac{1}{2} (\ket{0000} + \ket{1100} + \ket{0011} - \ket{1111})$.

\section{Continuations and Quantum Circuits}
\label{sec:four}

We now codify the analysis of the examples in the previous section in an
interpreter for the idealized quantum circuit language consisting of the \ccx\
and \had\ gates, described in~\cref{sec:eval}. In~\cref{sec:list}, we use the
interpreter to run three examples including the running example
of~\cref{fig:simon}.

\subsection{Continuation-based Evaluator}
\label{sec:eval}

The \ccx\ gate is a classical gate whose semantics is just as simple as the
semantics of $+$ or $*$. The Hadamard \had\ gate is the one interesting gate
that introduces choices. Specifically the application of \had\ introduces a
choice between 0 and $\pm 1$ with the proper weights. The evaluator
in~\cref{fig:eval} processes just one gate: it uses the delimited continuation
primitive \rkt{shift} to generate continuations, and a parameterized way to
\emph{collect} and process these continuations.

\begin{figure}[h]
  \begin{center}
    \begin{minted}{racket}
(define (evalg^ v g)
  (match `(,v ,g)
    ;; ccx with both control bits set
    [`((,d ,bs) (ccx ,ctrl1 ,ctrl2 ,targ))
     #:when (and (is-set? bs ctrl1)
                 (is-set? bs ctrl2))
     `(,d ,(neg bs targ))]
    ;; ccx with control bit(s) unset
    [`((,d ,bs) (ccx ,ctrl1 ,ctrl2 ,targ))
     `(,d ,bs)]
    ;; hadamard with target bit set
    [`((,d ,bs) (h ,targ))
     #:when (is-set? bs targ)
     (collect^ `(,(* hscale d) ,(neg bs targ))
               `(,(* -1.0 hscale d) ,bs))]
    ;; hadamard with target bit unset
    [`((,d ,bs) (h ,targ))
     (collect^ `(,(* hscale d) ,bs)
               `(,(* hscale d) ,(neg bs targ)))]))
\end{minted}
    \caption{\label{fig:eval}Continuation-based evaluator for quantum gates}
  \end{center}
\end{figure}

The \h{evalg^} function takes two arguments: \h{v} is a tuple of the form
\h{`(,d ,bs)}, where \h{d} is the probability amplitude, \h{bs} is a bit-vector
representing the state, and \h{g} is the quantum gate we are evaluating. The
first two clauses define the semantics of the \ccx\ gate by simply negating the
target qubit when the control qubits are set. If no control qubits are set, the
state is unchanged. The semantics of \had\ parallels the semantics of \h{choice}
in the previous section: the rest of the computation is duplicated, executed
once for each choice, and the choices are collected using a \h{collect^}
function. If the target qubit is set, we negate the amplitude, multiplying it by
\h{-1.0}.

\begin{figure}[h]
  \begin{center}
    \begin{minted}{racket}
(define (evalc^ v c)
  (foldl (lambda (g st) (evalg^ st g)) v c))

(define (run-circ t circ st)
  (reset (match (evalc^ st circ)
           [`(,d ,bs)
            (match t
              [`list (list `(,bs ,d))]
              ...)))))
\end{minted}
    \caption{\label{fig:evalc}Continuation-based evaluator for complete circuits}
  \end{center}
\end{figure}

As we represent circuits as lists of gates, to evaluate a circuit we simply fold
over the circuit, as shown in~\cref{fig:evalc}. Finally, the entry point to run
a circuit is to call \h{evalc^} with the circuit and the initial state,
enclosing the entire evaluation using a \h{reset} combinator which acts as the
delimiter for the captured continuation. We use an additional argument \h{t} in
that entry point which is a tag that lets us choose which instance of
\h{collect^} to use.

\subsection{Examples with List Collector}
\label{sec:list}

The interpreter is parameterized by an implementation of the function
\h{collect^} which dictates how we combine and manage the results of the
intermediate computations. We will explore two instances of \h{collect^}, one in
this section, and one in the next. In both instances, we first use \h{shift} to
capture the continuation \h{k}, and invoke it on the two arguments \h{x} and
\h{y} giving us access to the underlying states \h{a} and \h{b}. In this
section, we use the simplest strategy which collects all the states in a list:

\begin{minted}{racket}
(define (collect^ x y)
  (shift
   k
   (let ([a (k x)]
         [b (k y)])
     (cond
       ((and (list? a) (list? b)) (append a b))
       ...
\end{minted}

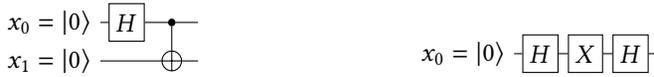
\begin{figure}[t]
  \centering
  \begin{tikzpicture}[scale=1.0]
    \begin{yquant*}
      qubit {$x_0 = \ket{0}$} x0;
      qubit {$x_1 = \ket{0}$} x1;
      h x0;
      cnot x1 | x0;
      align -;
    \end{yquant*}
  \end{tikzpicture}
  \qquad\qquad\qquad\qquad
  \begin{tikzpicture}[scale=1.0]
    \begin{yquant*}
      qubit {$x_0 = \ket{0}$} x0;
      h x0;
      x x0;
      h x0;
    \end{yquant*}
  \end{tikzpicture}
  \caption{\label{fig:qc}Two small quantum circuits}
\end{figure}

As examples of running circuits using this implementation, we consider two small
circuits, shown in~\cref{fig:qc}. The first circuit is a \had\ gate followed by
a \cx\ gate, operating on two qubits initialized to $\ket{00}$. We can represent
and execute it as follows, using lists for state management.

\begin{minted}{racket}
(run-circ `list
  (list (H 0)
        (CX 0 1))
  `(1.0 ,(make-vector 2 #f)))
\end{minted}
The execution happens as follows. The initial state is $\ket{00}$. At the \had\
gate, the rest of the computation, i.e., the application of \cx\ is duplicated
and executed once with the state $\frac{1}{\sqrt{2}} \ket{00}$ and once with the
state $\frac{1}{\sqrt{2}} \ket{10}$. The results are accumulated in a list
producing the final result~$[\frac{1}{\sqrt{2}} \ket{00}, \frac{1}{\sqrt{2}} \ket{11}]$.

The second circuit on the right in~\cref{fig:qc} is more interesting. It
consists of an~\had,~\xgate,~and~\had\ gate in series, on one qubit. We
represent and execute it as follows.

\begin{minted}{racket}
(run-circ `list
  (list (H 0)
        (X 0)
        (H 0))
  `(1.0 ,(vector #f)))  
\end{minted}
The first occurrence of \had\ duplicates the continuation consisting of the
application of \xgate\ followed by \had. That continuation itself generates a
nested choice, producing four distinct continuations in total. Tracing through
the evaluation of the circuit, the final answer is the list
$[\frac{1}{2} \ket{0}, -\frac{1}{2} \ket{1}, \frac{1}{2} \ket{0}, \frac{1}{2} \ket{1}]$.

However, this evaluation does not combine the probabilities, which is the only
physically realizable observation. The state $\ket{0}$ has two positive
probability amplitudes, which leads to constructive interference, and the state
$\ket{1}$ has two opposite amplitudes, which cancel each other out, leading to
destructive interference. Our current evaluator does not observe this, but we
can change that by switching to another instance of \h{collect^} in the next
section.

At this point, our main thesis becomes evident: quantum computing efficiently
(and somewhat mysteriously) manages these four distinct continuations using
constructive and destructive interference ensuring that the only possible
observable result is $\ket{0}$.

Before concluding this section, we confirm that evaluating the pure gates of the
circuit in~\cref{fig:simon} indeed produces the leaves of the tree shown
in~\cref{fig:tree}:

\begin{minted}{racket}
> (run-simon-list)
(+0.25|0000⟩)
(+0.25|0100⟩)
(+0.25|1000⟩)
(+0.25|1100⟩)
(+0.25|0011⟩)
(-0.25|0111⟩)
(+0.25|1011⟩)
(-0.25|1111⟩)
(+0.25|0011⟩)
(+0.25|0111⟩)
(-0.25|1011⟩)
(-0.25|1111⟩)
(+0.25|0000⟩)
(-0.25|0100⟩)
(-0.25|1000⟩)
(+0.25|1100⟩)
\end{minted}

\section{Continuations and Interference of Probability Amplitudes}
\label{sec:five}

In the previous section, we evaluated the pure part of a quantum circuit,
producing a list of states in the end. However, this evaluation does not combine
the probabilities, that is, the states at the end are in superposition, but the
probability amplitudes are not combined. By using lists to manage the quantum
states, we simply explored all branches of the search tree.

For example, consider the evaluation of the second circuit in~\cref{fig:qc}. The
final state is given by the list $[\frac{1}{2} \ket{0}, -\frac{1}{2} \ket{1},
\frac{1}{2} \ket{0}, \frac{1}{2} \ket{1}]$. Here, the state $\ket{0}$ has two
positive probability amplitudes, which leads to constructive interference, and
the state $\ket{1}$ has two amplitudes of opposing signs, which cancel each
other out, leading to destructive interference.

To allow our evaluator to combine the probabilities, we will use hash tables to
represent these states along with their probability amplitudes. The hash table
maps bit-vectors to probability amplitudes. All we have to do is change the
implementation of our \h{collect^} function.

\begin{minted}{racket}
  (define (collect^ x y)
    (shift
     k
     (let ([a (k x)]
           [b (k y)])
       (cond
         ...
         ((and (hash? a) (hash? b)) (hash-union a b #:combine +))))))
\end{minted}
To combine two hash tables, we take their union. If a bit-vector is repeated, we
simply add the probability amplitudes, using the \h{+} function to combine the
two values that are mapped to by the same key. This means that two branches with
the same state but opposite amplitudes will cancel each other out, and states
with amplitudes of the same sign will reinforce each other.

The runner initializes the hash table with the initial bit-vector and amplitude.
To run the example with the hash implementation, we simply change the tag passed
to it as an argument.

\begin{minted}{racket}
(define (run-circ t circ st)
  (reset (match (evalc^ st circ)
           [`(,d ,bs)
            (match t
              ...
              [`hash (hash bs d)])])))

(run-circ `hash
  (list (H 0)
        (X 0)
        (H 0))
  `(1.0 ,(vector #f)))
\end{minted}
The final observation now is $[1\ \ket{0}]$, as expected.

We also evaluate the circuit in~\cref{fig:simon} using the hash table
implementation. Following the evaluation tree in~\cref{fig:tree}, the states
connected by the dashed red arrows get annihilated, and the ones connected by
the solid blue arrows get reinforced.

\begin{minted}{racket}
> (run-simon-hash)
  (+0.50|0000⟩)
  (+0.50|1100⟩)
  (+0.50|0011⟩)
  (-0.50|1111⟩)
\end{minted}

\section{Reflections \& Conclusions}
\label{sec:conc}

We represented quantum states as bit-vectors with their probability amplitudes,
using lists and hash tables to store and combine them. Another way of
representing them is to directly work with probability distributions, which can
be approximately implemented using continuations. We quickly discuss a short
implementation of classical probability distributions using continuations, where
events are encoded as functions \rkt{(-> a p)}, and probability distributions
are encoded as higher-order functionals \rkt{(-> (-> a p) p)}.

We can write a \rkt{choose-p} function that takes two probability distributions
\rkt{k1} and \rkt{k2}, with a sampling bias \rkt{p}, and returns a new
distribution by calculating the convex combination of probabilities of events in
the two distributions.

\begin{minted}{racket}
(define (choose-p p k1 k2)
  (lambda (f)
    (let ([p1 (* p (k1 f))]
          [p2 (* (- 1.0 p) (k2 f))])
      (+ p1 p2))))
\end{minted}

To produce a constant probability distribution, given a value, for any event, we
apply the value to the event. The expectation of an event is computed by simply
applying the event to the underlying continuation.

\begin{minted}{racket}
(define (const-p a)
  (lambda (f) (f a)))

(define (expectation f k)
  (k f))
\end{minted}

Using these combinators, we can write yet another implementation of the
evaluator, where we use the \h{choose-p} operation (with no bias) to collect two
states. For example, we can execute the example in~\cref{fig:simon}.

\begin{minted}{racket}
> (run-simon-prob)
  (+0.25|0000⟩)
  (+0.25|0010⟩)
  (+0.25|0001⟩)
  (+0.25|0011⟩)
  (+0.25|1100⟩)
  (-0.25|1110⟩)
  (+0.25|1101⟩)
  (-0.25|1111⟩)
  (+0.25|1100⟩)
  (+0.25|1110⟩)
  (-0.25|1101⟩)
  (-0.25|1111⟩)
  (+0.25|0000⟩)
  (-0.25|0010⟩)
  (-0.25|0001⟩)
  (+0.25|0011⟩)
\end{minted}

However, it is obvious that these probability amplitudes do not have
cancellation, and we do not get the desired observation with interference. This
encoding of probability distributions only works for classical probabilities and
not quantum probabilities -- to move to quantum computing, we need to move from
classical probability theory to generalized probabilistic theory (GPT). We hope
to investigate these ideas connecting them to this work on continuations in the
future.

By using hash tables to represent probability distributions, we are sampling
from the distribution at every step of the computation when we evaluate a \had\
gate. On the other hand, by using continuations to represent probability
distributions, we're avoiding sampling until the very end, when a measurement is
made. Combining these techniques could lead to techniques for implementing
speculative execution in quantum computers.

We have presented a continuation-based analysis of quantum computing which shows
that it efficiently (and somewhat mysteriously) manages an exponential number of
continuations. It would be however incorrect to conclude that quantum computing
can manage arbitrary such collections of continuations as this would imply it
can solve NP-complete problems efficiently which is not believed to be true.
Discovering instances in which \h{collect} can be implemented without
necessarily invoking each continuation might provide insights on the elusive
power of quantum computing.

From a different perspective, it is known that there is a duality between values
and
continuations~\cite{Filinski89declarativecontinuations,10.1145/351240.351262}
with values flowing from producers to consumers and continuations flowing from
consumers to producers. This suggests a re-interpretation of our
continuation-based semantics as a distributed system with positive offers by
producers and negative counter-offers by consumers, which would be worth
investigating in detail.

\bibliography{cites.bib}

\appendix
\section*{Supplementary material}

We include the complete implementation of the evaluator and the circuit examples
discussed in~\cref{sec:four,sec:five}.

\begin{minted}{racket}
#lang racket

(require racket/control racket/hash racket/vector)

;; vector and hash helper operations

(define (upd v i f)
  (let ([w (vector-copy v)])
    (vector-set! w i (f (vector-ref w i)))
    w))

(define (neg v i)
  (upd v i not))

(define hscale
  (/ 1.0 (sqrt 2.0)))

(define (is-set? v i)
  (cond ((nonnegative-integer? i) (vector-ref v i))
        ((boolean? i) i)
        (else (error "is-set?: invalid index"))))

;; syntax for gates

(define-syntax-rule (X a) (CCX #t #t a))
(define-syntax-rule (CX a b) (CCX #t a b))
(define-syntax-rule (CCX a b c) `(ccx ,a ,b ,c))
(define-syntax-rule (H a) `(h ,a))

;; collect^ operation

(define (collect^ x y)
  (shift
   k
   (let ([a (k x)]
         [b (k y)])
     (cond
       ((and (list? a) (list? b)) (append a b))
       ((and (hash? a) (hash? b)) (hash-union a b #:combine +))
       (else (error "collect^: unknown state"))))))

;; gate and circuit evaluator

(define (evalg^ v g)
  (match `(,v ,g)
    [`((,d ,bs) (ccx ,ctrl1 ,ctrl2 ,targ))
     #:when (and (is-set? bs ctrl1)
                 (is-set? bs ctrl2))
     `(,d ,(neg bs targ))]
    [`((,d ,bs) (ccx ,ctrl1 ,ctrl2 ,targ))
     `(,d ,bs)]
    [`((,d ,bs) (h ,targ))
     #:when (is-set? bs targ)
     (collect^ `(,(* hscale d) ,(neg bs targ))
               `(,(* -1.0 hscale d) ,bs))]
    [`((,d ,bs) (h ,targ))
     (collect^ `(,(* hscale d) ,bs)
               `(,(* hscale d) ,(neg bs targ)))]
    [_ (error "evalg^: invalid arguments")]))

(define (evalc^ v c)
  (foldl (lambda (g st) (evalg^ st g)) v c))

;; pretty printer

(define (pretty-vec bs)
  (string-append
   "|"
   (foldl (lambda (b s) (string-append (if b "1" "0") s))
          "⟩"
          (vector->list bs))))

(define (pretty-prob d)
  (~r d #:sign '+ #:precision '(= 2)))

(define (pretty-state st)
  (cond
    [(list? st)
     (for-each
      (match-lambda*
        [`((,bs ,d)) #:when (< (abs d) 0.01) (void)]
        [`((,bs ,d)) (printf "(~a~a)~n" (pretty-prob d) (pretty-vec bs))])
      st)]
    [(hash? st)
     (hash-for-each
      st
      (match-lambda*
        [`(,bs ,d) #:when (< (abs d) 0.01) (void)]
        [`(,bs ,d) (printf "(~a~a)~n" (pretty-prob d) (pretty-vec bs))]))]
    [else (error "pretty-state: unknown state")]))

;; runner

(define (run-circ t circ st)
  (reset (match (evalc^ st circ)
           [`(,d ,bs)
            (match t
              [`list (list `(,bs ,d))]
              [`hash (hash bs d)]
              [_ (error "run-circ: invalid tag")])])))

;; H-CX example

(define (run-hcx-list)
  (pretty-state (run-circ `list
                          (list (H 0)
                                (CX 0 1))
                          `(1.0 ,(make-vector 2 #f)))))

(define (run-hcx-hash)
  (pretty-state (run-circ `hash
                          (list (H 0)
                                (CX 0 1))
                          `(1.0 ,(make-vector 2 #f)))))

;; H-X-H example

(define (run-hxh-list)
  (pretty-state (run-circ `list
                          (list (H 0)
                                (X 0)
                                (H 0))
                          `(1.0 ,(vector #f)))))

(define (run-hxh-hash)
  (pretty-state (run-circ `hash
                          (list (H 0)
                                (X 0)
                                (H 0))
                          `(1.0 ,(vector #f)))))

;; Simon example

(define (run-simon-list)
  (pretty-state (run-circ `list
                          (list (H 0)
                                (H 1)
                                (CX 0 2)
                                (CX 0 3)
                                (CX 1 2)
                                (CX 1 3)
                                (H 0)
                                (H 1))
                          `(1.0 ,(make-vector 4 #f)))))

(define (run-simon-hash)
  (pretty-state (run-circ `hash
                          (list (H 0)
                                (H 1)
                                (CX 0 2)
                                (CX 0 3)
                                (CX 1 2)
                                (CX 1 3)
                                (H 0)
                                (H 1))
                          `(1.0 ,(make-vector 4 #f)))))


;; swap-test using Fredkin gate
(define (run-swap-test-hash)
  (pretty-state (run-circ `hash
                          (list (H 0)
                                (CCX 0 1 2)
                                (CCX 0 2 1)
                                (CCX 0 1 2)
                                (H 0))
                          `(1.0 ,(vector #f #f #f)))))
\end{minted}

\end{document}